\documentclass[prl,twocolumn,showpacs,showkeys,amsfonts,amssymb]{revtex4}
\usepackage{amsmath}
\usepackage{graphics}
\usepackage[pdftex]{graphicx}
\usepackage{graphicx}
\usepackage{dcolumn}
\usepackage[usenames,dvipsnames]{color}
\usepackage[normalem]{ulem}

\begin{document}
\title{Using a P\'{e}clet Number for the Translocation of a Polymer through a Nanopore to Tune Coarse-Grained Simulations to Experimental Conditions}
\author{Hendrick W. de Haan}
\affiliation{Faculty of Science, University of Ontario Institute of Technology, Oshawa, Ontario, Canada, L1H 7K4}

\author{David Sean}
\author{Gary W. Slater}
\affiliation{Physics Department, University of Ottawa, Ottawa, Ontario, Canada, K1N 6N5}

\date{\today}
\begin{abstract}
Coarse-grained simulations are often employed to study the translocation of DNA through a nanopore.
The majority of these studies investigate the translocation process in a relatively generic sense and do not endeavour to match any particular set of experimental conditions.
In this manuscript, we use the concept of a P\'{e}clet number for translocation, $P_t$, to compare the drift-diffusion balance in a typical experiment \textit{vs} a typical simulation.
We find that the standard coarse-grained approach over-estimates diffusion effects by anywhere from a factor of 5 to 50 compared to experimental conditions using dsDNA.
By defining a coarse-graining parameter, $\lambda$, we are able to correct this and tune the simulations to replicate the experimental $P_t$ (for dsDNA and other scenarios).
To show the effect that a particular $P_t$ can have on the dynamics of translocation,
we perform simulations across a wide range of $P_t$ values for two different types of driving forces: a force applied in the pore and a pulling force applied to the end of the polymer.
As $P_t$ brings the system from a diffusion dominated to a drift dominated regime, a variety of effects are observed including
a non-monotonic dependence of the translocation time $\tau$ on $P_t$ and a steep rise in the probability of translocating.
Comparing the two force cases illustrates the impact of the crowding effects that occur on the $trans$ side:
a non-monotonic dependence of the width of the $\tau$ distributions is obtained for the in-pore force but not for the pulling force.
\end{abstract}
\maketitle

\section{Introduction}

The translocation of a polymer through a nanopore has been the subject of intense study in recent years,
both due to its biological relevance as well as emerging nanotechnology applications such as nanopore-based devices for detecting and sequencing DNA.
Among this body of work, there have been numerous Molecular Dynamics simulation studies employing a coarse-grained (CG) approach in which a relatively generic polymer is simulated.
As shown in Table \ref{paramsTable}, a typical CG setup consists of a polymer on the order of 100 monomers ($N=100$), an external driving force of $F=1$--$10~\epsilon/\sigma$ (where $\sigma$ is the monomer or bead diameter, and $\epsilon$ is the characteristic energy, typically the strength of the bead-bead interaction), and a thermal energy of $k_\mathrm{B}T \approx \epsilon$.
In this paper, we consider the physical appropriateness of such setups for modelling the translocation of polyelectrolytes such as double stranded DNA (dsDNA) . We thus consider a typical setup of $N=100$, $F \approx \epsilon/\sigma$, $k_\mathrm{B}T \approx \epsilon$ and investigate what experimental conditions this corresponds to.

\begin{table}[]
\begin{center}
\begin{ruledtabular}
\begin{tabular}{cccc}
Reference& \ \ \ \ \ \ \ \  N  \ \ \ \ \ \ \ \  & $F$ [$\frac{\epsilon}{\sigma}$] & \ \ \ \ $k_\mathrm{B}T$ [$\epsilon$] \ \ \ \ \\
\hline
\cite{dubbeldam2012forced}	&100--500 & 1--20	&	1.2\\
\cite{lehtola2008}	&	100--400	&	10	&	1\\
\cite{lehtola2010pore}	&	256	&	0.05--5	&	1.2\\
\cite{linna2012event}	&	200	&	0--2	&	1\\
\cite{luo2009driven}	&	16--256*	&	0.5--10	&	1.2\\
\cite{dubbeldam2013driven}	&	100	&	0--10	&	1\\
\cite{bhattacharya2009scaling}	&	8--256	&	4--6	&	1.5\\
\cite{lehtola2010unforced}	&	31--251	&	0.3--10	&	1\\
\cite{lehtola2009dynamics}	&	100	&	1--100	&	1\\
\end{tabular}
\end{ruledtabular}
\end{center}
\caption{Values of the degree of polymerization $N$, the driving force $F$ and the thermal energy $k_\mathrm{B}T$ found in the literature for 3D Langevin Dynamics simulations of translocation (* indicates estimated values).
\label{paramsTable}}
\end{table}%
To do so, we define a P\'{e}clet number for translocation, $P_t$, such that we can quantify the balance between the drift (external force) and diffusive (thermal noise) components of the translocation process.
We then introduce a coarse-grained scaling factor $\lambda$ such that allows us to easily change this balance.
Low $\lambda$ values correspond to a diffusion dominated process while high $\lambda$ values yield a drift dominated process.

To test the role of the diffusion-drift balance, we simulate polymer translocation across a wide range of $P_t$ values.
We demonstrate that the translocation process changes significantly as we move from a diffusive to a drift dominated regime. 
Effects include a steep rise in the probability of translocation,
a non-monotonic increase in the translocation time,
and a non-monotonic decrease in the spread of translocation times.

At the end of the manuscript, we consider different experimental translocation scenarios: dsDNA, ssDNA and rod-like viruses.
These examples demonstrate the usefulness of using $P_t$ to match simulations to experimental conditions via measurable quantities.
By estimating the $\lambda$ value appropriate for each scenario, we also show that the standard approach is unphysical for certain cases.
Most dramatically, we find that typical CG setups for dsDNA over-estimate diffusive effects by a factor of 5-50.

\section{Theory}

\subsection{P\'{e}clet number for translocation}

As for most transport-based processes, the translocation of a polymer through a nanopore contains aspects of both diffusive and driven dynamics, the latter arising primarily from the presence of external forces. The balance between these two mechanisms can greatly affect the dynamics of translocation. In the limit of zero field, or unbiased translocation, 
a polymer started half-way through the pore will exhibit stochastic, random-walk behaviour (albeit sub-diffusive) for the majority of the process. On the other hand, in the limit of zero diffusivity, the trajectory of a given translocating polymer will be deterministic.

To characterize the balance between drift and diffusion, 
we define a dimensionless parameter given by the ratio of the free solution polymer relaxation time $\tau_\mathrm{relax}$ over the translocation time $\tau_\mathrm{trans}$.
As it characterizes the drift-diffusion balance, this metric is essentially a P\'{e}clet number \cite{squires2005} for translocation:
\begin{equation}
P_t = \frac{\tau_\mathrm{relax}}{\tau_\mathrm{trans}}.
\end{equation}
A similar argument was given by Saito and Sakaue \cite{Saito2012} as well as Dubbeldam et al. \cite{dubbeldam2013driven}
where a P\'{e}clet  number was given in terms of simulation parameters such as the friction coefficient.
Our goal here is to match coarse-grained simulations to experimental conditions and thus we define $P_t$ in terms of quantities obtainable both experimentally and \textit{in silico}.
As usual, the free solution-relaxation time is given by the expression:
\begin{equation}
\tau_\mathrm{relax} = \frac{R_{go}^2}{D_o},
\end{equation}
where $R_{go}$ is the equilibrium radius of gyration and $D_o$ is the free solution diffusion coefficient.
Our P\'{e}clet number is then given in terms of three observables:
\begin{equation}
P_t = \frac{1}{\tau_\mathrm{trans}} \frac{R_{go}^2}{D_o}.
\label{good_pec}
\end{equation}
In order to match experimental and simulation results,
we will now develop quantitative expressions for $P_t$ for both cases.
Starting with the latter, it is first necessary to introduce our simulation approach.

\subsection{Simulation Setup}

\subsubsection{Langevin Dynamics Simulations}

We employ a coarse-grained simulation approach that is very common in molecular-dynamics based translocation work \cite{dubbeldam2012forced, lehtola2008, lehtola2010pore, linna2012event, luo2009driven, dubbeldam2013driven, bhattacharya2009scaling, lehtola2010unforced, lehtola2009dynamics, slater2009}.
Performing Langevin Dynamics (LD), the effects of the solvent are included implicitly and the resulting equation of motion is
\begin{equation}
m \dot{\vec{v}} = \vec{F} - \vec{\nabla} U (\vec{r}) - \gamma \vec{v} + \sqrt{ 2 \gamma k_\mathrm{B} T }~\vec{\xi}(t)
\label{eomLD}
\end{equation}
where $m$ is the mass of the particle, $\vec{v}$ is the velocity, $\vec{F}$ is the applied external force, $-\vec{\nabla} U(\vec{r})$ is the sum of the conservative forces, $\gamma$ is the friction coefficient and the term $-\gamma \vec{v}$ represents the damping effects of the fluid. The last term in Eq. \eqref{eomLD}, in which $\vec{\xi}$ \cite{slater2009} is a random vector, models the random kicks of the solvent.

From Eq. \eqref{eomLD}, it is straightforward to show that for a free particle ($\nabla U(\vec{r})=0$), the diffusion coefficient $D$ and drift velocity $v_\mathrm{drift}$ are given by
\begin{equation}
D = \frac{k_\mathrm{B} T}{\gamma},
\label{reg_D}
\end{equation}
and
\begin{align}
\vec{v}_\mathrm{drift} & = \frac{\vec{F}}{\gamma}.
\label{reg_av}
\end{align}
Finally, using the equipartition theorem, the thermal velocity of the particle is given by:
\begin{equation}
v_\mathrm{th} = \sqrt{\frac{3 k_\mathrm{B}T}{m}}.
\label{equi}
\end{equation}

\subsubsection{Coarse-Graining}

Equations \ref{reg_D}, \ref{reg_av}, and \ref{equi} characterize the dynamics of a single particle.
In a typical coarse graining procedure, we take a single simulation bead as a model for many individual particles of mass $m$.
Consider such a bead to represent $\lambda$ smaller units.
In the case of free-draining hydrodynamics, each particle contributes independently to the friction of the simulation bead such that its net friction is $\lambda \gamma$.
Similarly, the mass of the bead is $\lambda m$, and if each sub-particle feels a force $F$, the net force is $\lambda F$.
Hence, using the following transformations
\begin{eqnarray}
\gamma & \rightarrow & \lambda \gamma, \\
\vec{F} & \rightarrow & \lambda \vec{F}, \\
m & \rightarrow & \lambda m,
\end{eqnarray}
equations \ref{reg_D}, \ref{reg_av}, and \ref{equi} become
\begin{eqnarray}
D & = &  \frac{k_B T}{\lambda \gamma},\label{good_D} \\
\vec{v}_\mathrm{drift} & = & \frac{\lambda \vec{F}}{\lambda \gamma} = \frac{\vec{F}}{\gamma}, \label{good_av} \\
   v_\mathrm{th} & = & \sqrt{\frac{3 k_{\mathrm{B}}T}{\lambda m}}. \label{good_v}
\end{eqnarray}
The drift velocity is thus unaffected in the free-draining limit.
We will use $\lambda$ as a coarse-graining parameter (with this general definition, it is possible to use $\lambda < 1.0$). We will use the fact that the physics of translocation depends on $P_t$ and that $P_t$ itself is a direct function of $\lambda$ to use the latter parameter to tune the simulations to attain the experimentally relevant conditions.

\subsubsection{Polymer Simulations}
To model the forced translocation of a polymer chain, we use a common CG approach \cite{slater2009}.
A purely repulsive WCA potential is used for excluded volume interactions between beads \cite{weeks1978}:
\begin{equation}
U_\mathrm{WCA} (r) = \label{EQ:WCA}
\begin{cases}
8 \epsilon \left[ \left( \frac{\sigma}{r}\right)^{12} - \left( \frac{\sigma}{r} \right)^6 \right]   &\text{for } r <  2^{\frac{1}{6}}\sigma  \\
0 &\text{for } r \geq 2^{\frac{1}{6}}\sigma.
\end{cases}
\end{equation}
This defines the simulation units of length ($\sigma$, the size of the monomers) and energy ($\epsilon$). The WCA potential is also used to model the membrane as a continuous surface with a pore of radius 1 $\sigma$. This ensures single file translocation \cite{de2010mapping}. To link connected monomers, the FENE potential is used; similar to the work of Kremer and Grest, we set its maximum spring extension to $1.5 \sigma$ \cite{slater2009, grest1986}. For reasons which will be clarified later, stable interactions are needed between connected beads over an unconventionally wide range of thermal forces. To achieve this, we increased the prefactors for the WCA and FENE potentials by a factor of two from what is typically found. We thus use a FENE spring constant of $k=60$~$\epsilon / \sigma^2$.

As the probability of translocation is very low at low $\lambda$ values, we initiate the polymer with one monomer on the $trans$ side. We examine two different force scenarios: a driving force $F=1.0$~$\epsilon / \sigma$ applied to the monomer(s) located in the pore, and a pulling force $F=1.0$~$\epsilon / \sigma$ applied on the end monomer (see cartoon insets of Fig.~\ref{fig:distros}). From now on, we will drop the LD units for simplicity.

\subsection{The Simulation P\'{e}clet Number}

From the definition in Eq.~\ref{good_pec}, we require the free solution diffusion coefficient, the equilibrium radius of gyration, and the translocation time of a polymer of $N$ beads.
Since there is no hydrodynamic coupling in LD, the net friction coefficient scales like $D_o = k_\mathrm{B}T / \gamma N$. Incorporating the coarse-graining $\lambda$ factor we obtain
\begin{equation}
D_o = \frac{k_\mathrm{B} T}{\lambda \gamma N}.
\end{equation} 
Simulations were performed at $k_\mathrm{B}T=\epsilon$ to determine $\tau_\mathrm{trans}$ as a function of $F$, $N$, and $\gamma$. Our data agree with 
\begin{equation}
\tau_\mathrm{trans} = A \frac{\gamma}{F} N^{\alpha}.
\end{equation}
For $F=1.0$, $\gamma=1.0$ , and $N=50,100, 200$,
$A$=3.28~[$\sigma$] and $\alpha$=1.43 is the effective scaling exponent, in good agreement with previous studies (\textit{e.g.}, \cite{bhattacharya2009scaling,luo2009driven}). These parameters are only weakly dependent on $k_\mathrm{B}T$ (data not shown). Finally, the equilibrium radius of gyration can be fitted using the standard expression 
\begin{equation}
R_{go} = B N^{\frac{3}{5}}
\end{equation}
where $B=0.468$~[$\sigma$] in our case. Using Eq. \ref{good_pec}, our simulation translocation P\'{e}clet number is thus given by
\begin{eqnarray}
P_\mathrm{sim} & \approxeq & \left( \frac{F}{A \gamma N^{\alpha}} \right)  \frac{  \left( B N^{\frac{3}{5}}\right) ^2}{ \frac{k_\mathrm{B} T}{\lambda \gamma N} } \\ 
   & \approxeq & \frac{C  \lambda F  N^{0.77}}{k_\mathrm{B}T }
\label{pec_sim}
\end{eqnarray}
where $C=0.067$~[$\sigma$] ($P_\mathrm{sim}$ is dimensionless as required). Note that $P_\mathrm{sim}$ is independent of the friction coefficient. 

The scaling $P_\mathrm{sim} \sim \lambda$ allows us to tune the simulation P\'{e}clet number via the coarse-graining parameter $\lambda$. In practice though, a LD simulation does not directly include $\lambda$; instead, $\lambda$ is used in an implicit way when interpreting the simulation data. However, Eq. \ref{pec_sim} shows that we can effectively tune $\lambda$ (as we shall see, we will need to increase $P_\mathrm{sim}$) by changing one (or any combination of) the three simulation parameters $N$, $F$, or $k_\mathrm{B}T$  as discussed below.
\begin{enumerate}
\item Using longer polymers by increasing $N$ will increase $P_\mathrm{sim}$, but this quickly becomes prohibitively expensive in terms of computation time.

\item While increasing $F$ works well for small increases in $F$, it generally becomes problematic at very large forces because the friction parameter is unchanged. Unless one is careful, very high forces may lead to situations where we are no longer in the overdamped regime that is intrinsic to nanofluidics. This limits our ability at tuning $P_\mathrm{sim}$ via $F$.
\item We can also change $P_\mathrm{sim}$ by varying the thermal factor $k_\mathrm{B}T$ in the simulations. This is clearly the best approach since Eq. \ref{good_D} and \ref{good_v} indicate that what actually matters is the ratio $k_\mathrm{B}T/\lambda$. Changing $1/k_\mathrm{B}T$ is thus fully equivalent to changing $\lambda$. 
\end{enumerate}

\section{Results}

\begin{figure}[h]
 	\centering
	\includegraphics[width=0.50\textwidth]{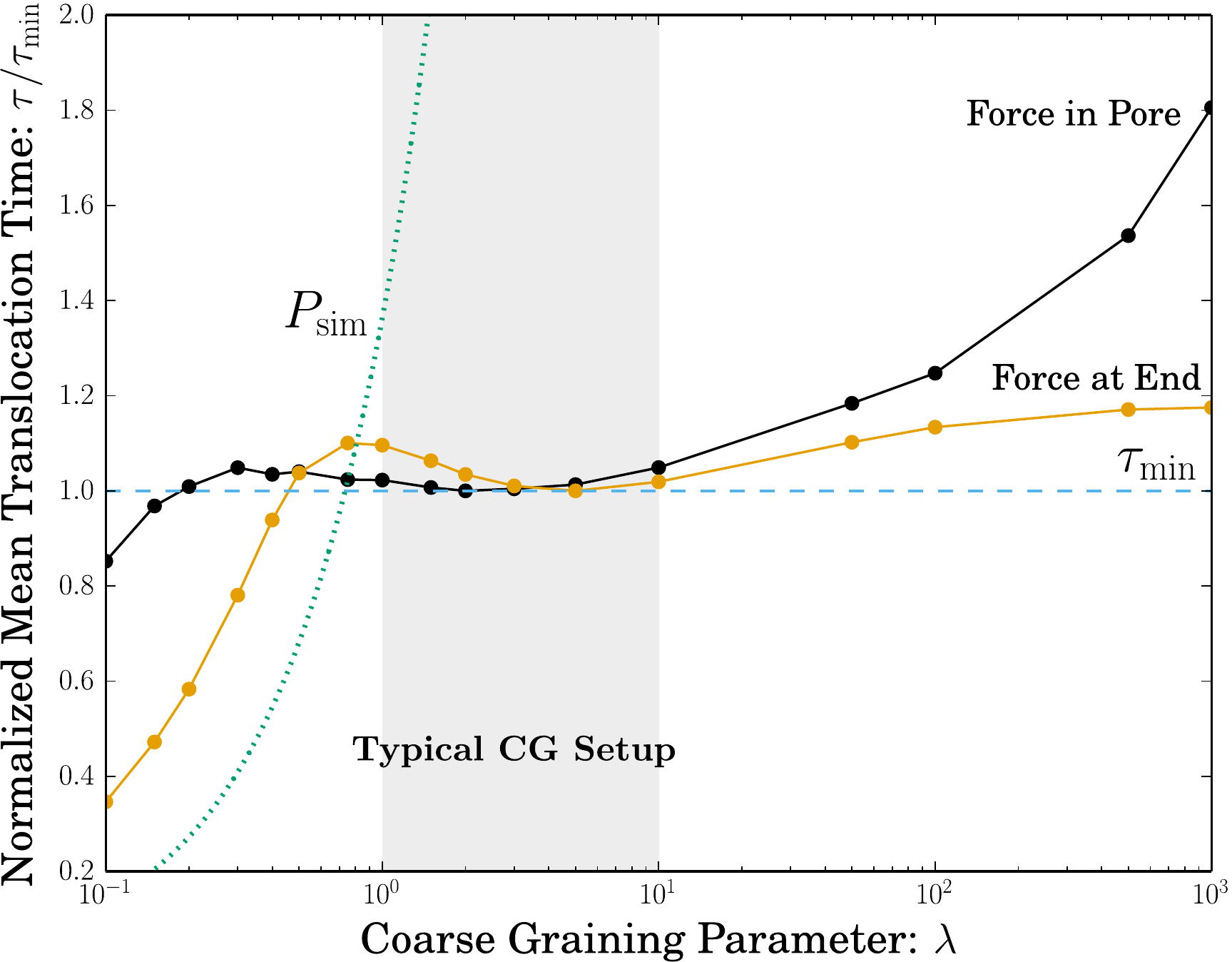} 
	\caption{Translocation time $\tau$ (normalized by the local minimum $\tau_\mathrm{min}$) vs. $\lambda$
	for the in-pore force case (black) and for pulling at the end case (orange) for a polymer chain of size $N=50$.
	The corresponding simulation P\'{e}clet number $P_\mathrm{sim}$ is shown as a dotted green line.
	\label{fig:tau}}
\end{figure}

To demonstrate how the P\'{e}clet number can affect translocation dynamics,
we have performed simulations of driven translocation across a wide range of $\lambda$.
Simulations were performed for polymers of length $N=50$ and $N=100$.
We focus on the $N=50$ results as it is easier to resolve all relevant regimes for the shorter polymer.
Two driven translocation cases were studied:
i) a force applied to any monomer that is in the pore and
ii) a force applied to the lead monomer which pulls the polymer through the pore.
The results for the translocation time are shown in Fig. \ref{fig:tau}.
In discussing the results, we present the data in terms of the dependence on the coarse-graining parameter, $\lambda$, where different values of $\lambda$ correspond to different P\'{e}clet numbers.

Non-monotonic results are obtained for both force scenarios.
From the results, we define three regimes.
At low $\lambda$ values (high T), $\tau$ increases with increasing $\lambda$.
For intermediate $\lambda$ values, $\tau$ decreases slightly with increasing $\lambda$.
And finally, at high $\lambda$ values (low T), $\tau$ again increases with increasing $\lambda$.
Given the definition of $P_t$, we can associate these three regimes with the relative drift-diffusion balance:
i) low $\lambda$ values correspond to a diffusion dominated regime
ii) intermediate $\lambda$ values correspond to a transition regime
and iii) high $\lambda$ values correspond to a primarily driven regime.

\begin{figure}[h]
 	\centering
	\includegraphics[width=0.50\textwidth]{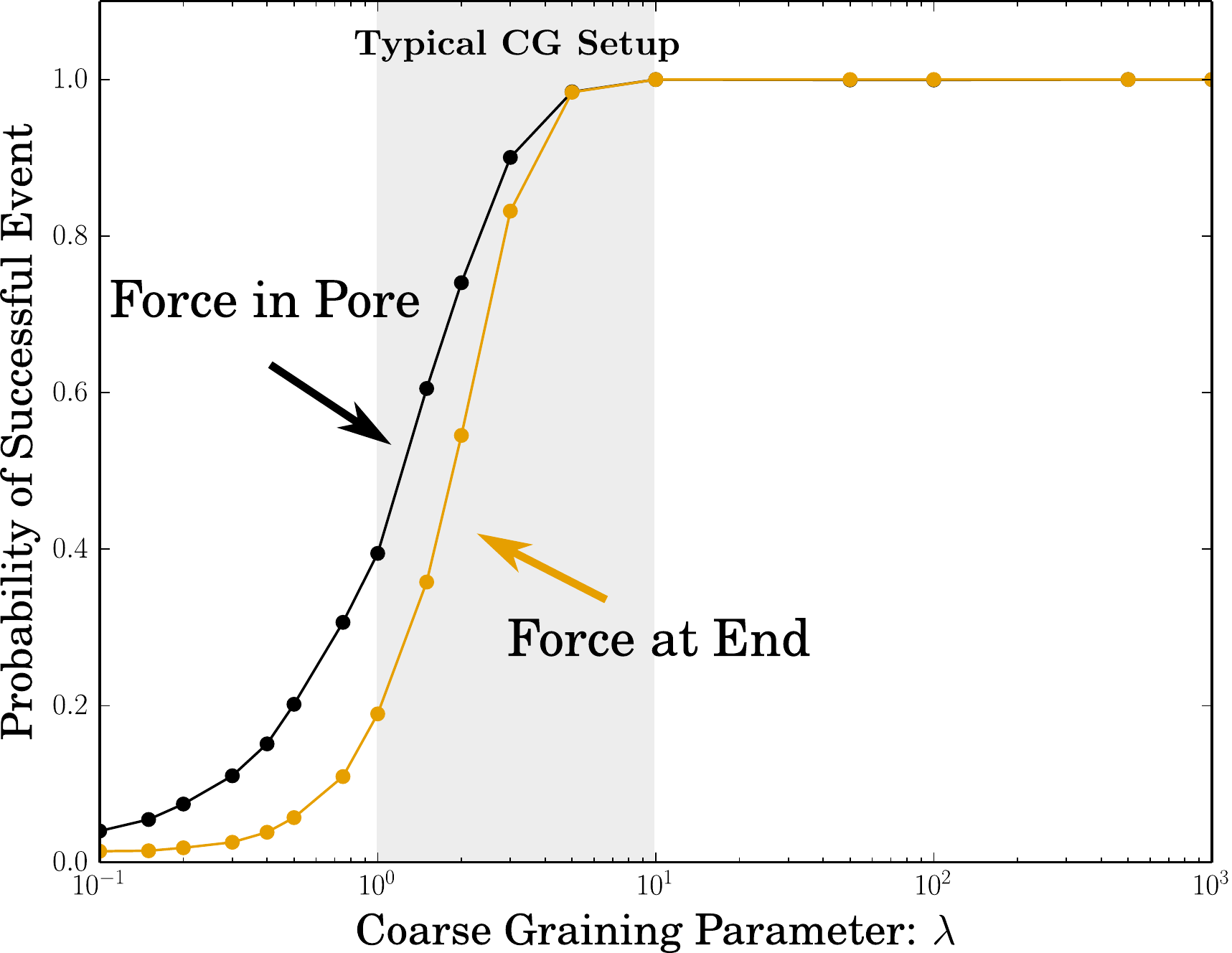} 
	\caption{Probability of a successful translocation event vs. $\lambda$ for the in-pore force (black) and force at end (orange) cases for a polymer chain of size $N=50$.
	\label{fig:prob}}
\end{figure}

These regimes can be verified by examining the probability of translocation as shown in Fig. \ref{fig:prob}.
At low $\lambda$, diffusion is dominant over drift and thus the polymer frequently retracts to the $cis$ side.
At high $\lambda$, the thermal noise is suppressed and thus, as the process becomes purely deterministic, the polymer never retracts and the translocation probability goes to 1.
In the transition regime, the probability rapidly increases between these two limits (note the $x$=axis is logarithmic).
Having outlined three regimes, we explore the physics of translocation in each one.

\subsection{Diffusive: Low $\lambda$}

At low $\lambda$, $\tau$ increases with increasing $\lambda$.
This regime corresponds to the quasi-static limit for translocation.
Given the high diffusivity, the relaxation time of the polymer is very short and in fact is shorter than the typical time for monomers to translocate through the pore.
Hence, the polymer is relaxed at each stage of translocation.
This allows for an approximation in which the polymer is reduced to a single particle between two absorbing walls  
subject to diffusion-drift while crossing an entropic barrier \cite{sung1996,muthu1999,hdh2011incremental}.
The validity of this approximation has been demonstrated in simulations where the high diffusivity is achieved by lowering the viscosity of the fluid \cite{dehaan2012,dehaan2013}.

In this regime, two factors contribute to $\tau$ decreasing as $\lambda$ decreases.
First, a higher diffusion coefficient means that the rate of motion for the polymer increases which result in faster translocation, and thus higher diffusion aids translocation.
More importantly, there is also a selection process implicit to the dynamics at low $\lambda$.
This can be seen in Fig.~\ref{fig:prob} where the probability of translocation is very low for $\lambda < 1.0$.
As $\lambda$ decreases, the probability of translocation decreases further.
This means that, translocation only occurs for those events in which the polymer quickly moves towards the $trans$ side;
i.e., only the fastest events survive and $\tau$ decreases.

\begin{figure}[h]
 	\centering
	\includegraphics[width=0.50\textwidth]{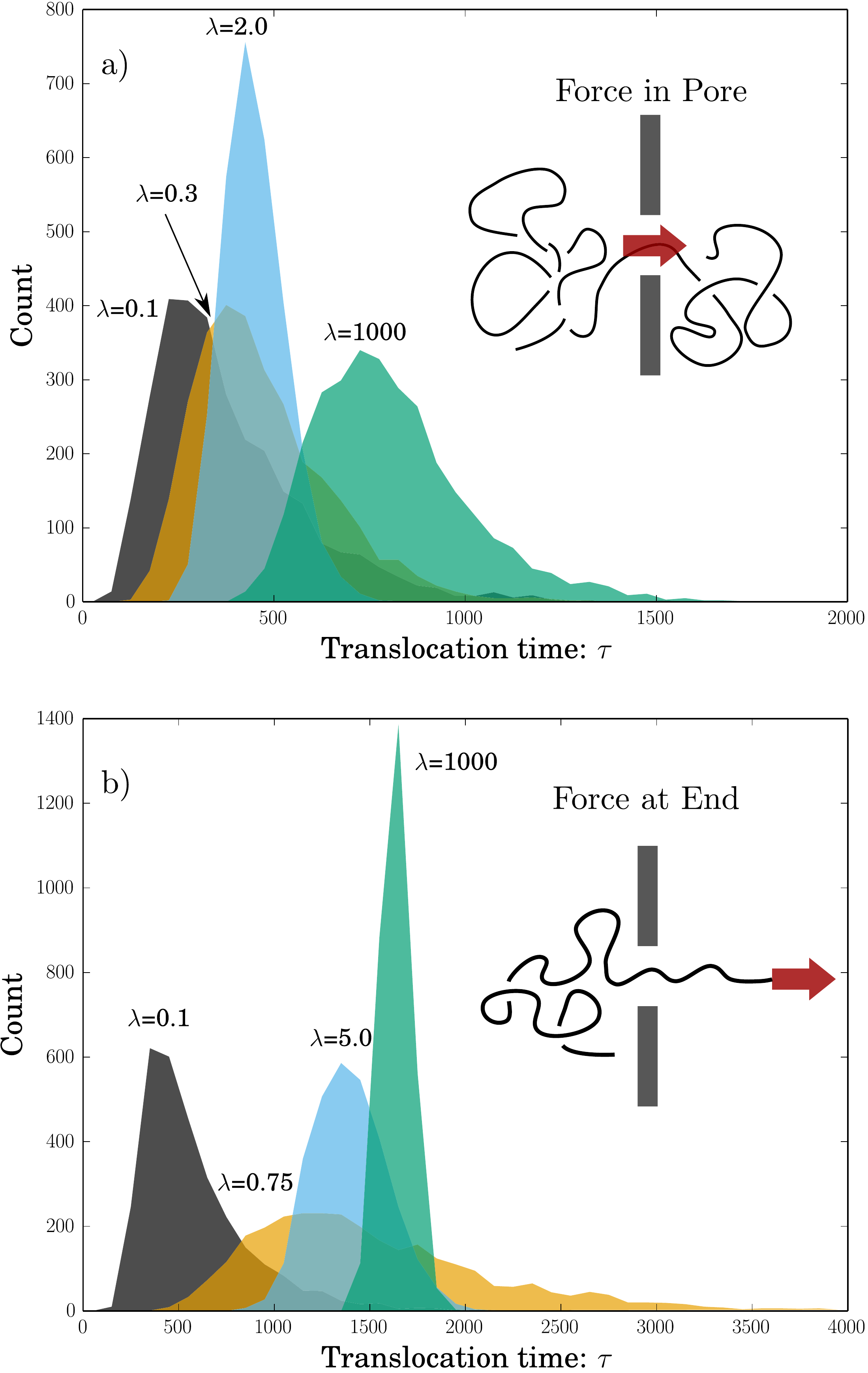} 
	\caption{Distribution of translocation times for the force in pore (top) and force at end (bottom) cases for a polymer chain of size $N=50$.
	In both cases, four different $\lambda$ values are shown.
	}
	\label{fig:distros}
\end{figure}

To examine the details of how $\tau$ increases with $\lambda$, the distributions for four selected $\lambda$ values are shown in Fig.~\ref{fig:distros}.
For both the in-pore and pulling forces, at low $\lambda$ values, the distributions shift to longer translocation times as $\lambda$ increases; this agrees with the physical picture outlined above.

\subsection{Transition: Intermediate $\lambda$}

Examining the distributions beyond the low $\lambda$ values, the nature of the transition region becomes clear.
Although the distributions continue to shift to larger $\tau$ values, the width of the distributions begins to decrease significantly.
As the dynamics transition from diffusion to drift, the contribution to the variance in $\tau$ arising from the thermal noise rapidly diminishes.
Most significantly, this reduction in the impact of the stochastic path reduces the instances of long-time events.
Hence, even though the most probable value of $\tau$ is increasing, the mean of $\tau$ is actually decreasing.
This yields the non-monotonic nature of the transition region.

\begin{figure}[h]
 	\centering
	\includegraphics[width=0.50\textwidth]{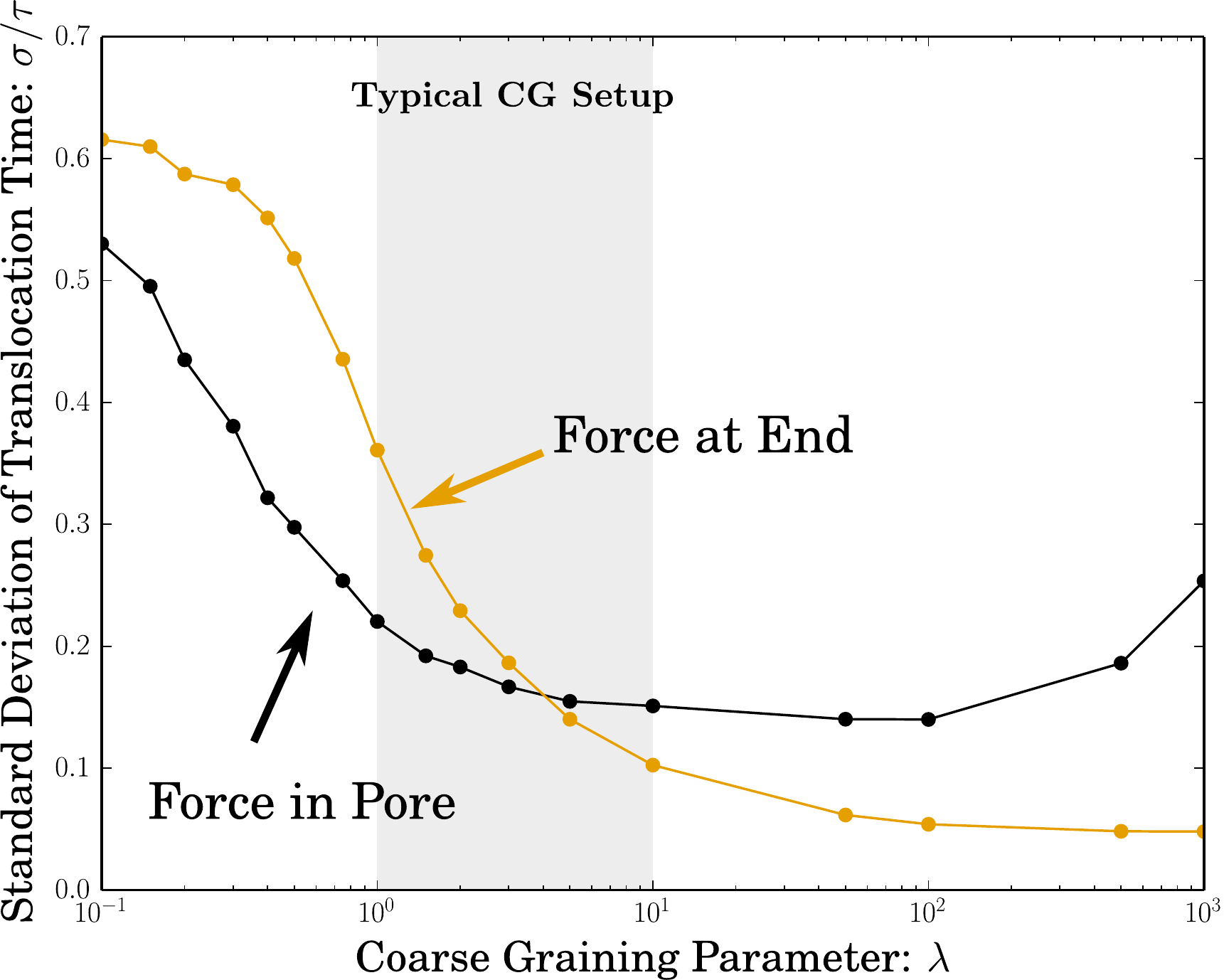} 
	\caption{Standard deviation of the translocation time distributions, $\sigma_{st}$ (normalized by the mean, $\tau$) vs. $\lambda$ for a polymer of size $N=50$.
	}
	\label{fig:std}
\end{figure}

To quantify this, the standard deviation, $\sigma_{st}$, normalized by the mean translocation time $\tau$ for all $\lambda$ values is shown in Fig. \ref{fig:std}.
For the pulling force, the normalized width continuously decreases with increasing $\lambda$.
For the in-pore force case, $\sigma_{st}/\tau$ decays with increasing $\lambda$ both in the diffusion and transition regimes, but it increases when $\lambda$ is very large.

\subsection{High $\lambda$: driven translocation}

For the pulling force case, the normalized standard deviation $\sigma_{st}/\tau$ at high $\lambda$ values appears to be plateauing at a finite value.
As shown in Fig. \ref{fig:tau}, $\tau$ plateaus at high $\lambda$ values.
If thermal factors are negligible, the dynamics are essentially fully deterministic, and hence increasing $\lambda$ has no further effect.
Recall there are two factors contributing to the variance in $\tau$ for the pulling case: variations in the stochastic path and in the initial conformation of the polymer \cite{Saito2012, lu2011origins}.
As $\lambda$ increases, the first factor becomes negligible but the second remains since it is independent of the $\lambda$ value.
Hence, both $\tau$ and $\sigma_{st}$ level off at high $\lambda$ and thus so does the normalized width $\sigma_{st}/\tau$.

The behaviour is very different for the in-pore force where $\sigma_{st}/\tau$ increases with $\lambda$ at high $\lambda$
(this increase can also be seen in the distribution plots, Fig.~\ref{fig:distros}).
This difference in the $\sigma_{st}/\tau$ plots highlights one of the main differences between the two force cases:
for the pulling force, there is no crowding on the \textit{trans} side.
For the in-pore force, once the monomers are through the pore, there is no force on them.
At high $\lambda$ values, they do not diffuse away quickly and thus significant crowding is observed.
This crowding presents a steric barrier to translocation: the translocation time for the in-pore force 
increases much more dramatically than for the pulling force where it is actually levelling off (Fig.~\ref{fig:tau}).

As $\tau$ increases, one expects $\sigma_{st}$ to increase as well since the stochastic path will have a larger impact.
However, $\sigma_{st}/\tau$ increases at high $\lambda$ indicating that $\sigma_{st}$ is increasing at a faster rate than $\tau$.
The reason for this is that crowding also presents an additional source of variation for the translocation time. 
The degree of crowding for any particular event will depend on the details of individual \textit{trans} conformations.
Hence, the translocation time is dependent not only on the initial conformation and the stochastic path, but also the \textit{trans} conformation.
This extra source of variance results in $\sigma_{st}/\tau$ increasing at high $\lambda$.
Again, this is dramatically different from the pulling case where only the first two factors are present and $\sigma_{st}/\tau$ levels off.

Considering the increase in $\tau$ in this force-in-pore regime, recall that the probability of translocation is $\approx 1.0 $.
Hence, the increase in $\tau$ with increasing $\lambda$ cannot be due to a selection process as it was in the low $\lambda$, diffusive regime. 
The increase can be seen by considering the relaxation of a \textit{cis} subchain that just lost monomers to translocation: the relaxed conformation of this shorter subchain has a centre-of-mass closer to the pore.
Hence relaxation on the \textit{cis}-side biases the remaining monomers towards the pore, aiding translocation.
As $\lambda$ increases, this relaxation is suppressed (the \textit{cis} subchain remains stretched away form the pore) and thus $\tau$ increases.

\section{Experimental Scenarios: Tuning $\lambda$}

In this section we examine several different experimental translocation scenarios.
For each case, we first estimate the experimental value of the PŽclet number and we then find how $\lambda$ can be used to tune the simulations appropriately.

\subsection{Double Stranded DNA}
Smith et al. \cite{smith1996dynamical} measured the diffusion coefficient for a wide range (4--309 kbp) of dsDNA sizes.
From their data we get
\begin{equation}
D_{o} \approx \frac{2.38}{L^{0.608}} \textrm{{$\mu$m}$^2$/s},
\end{equation} 
where $L$ is the polymer length in $\mu$m.
Smith et al. also extracted the radius of gyration from their data, from which we obtain
\begin{equation}
R_{go} = 0.146 L^{\frac{3}{5}} \textrm{$\mu$m}.
\end{equation}
For the translocation time, we use two seminal studies for dsDNA through solid state pores.
First, Storm et al. in 2001 studied the translocation of DNA strands 2.2-32.6 $\mu$m in length and found a scaling of $\tau \sim L^{1.27}$ \cite{storm2005fast}. 
From their data we estimate
\begin{equation}
\tau_\mathrm{trans} \approx 5.6 \times 10^{-5} L^{1.27}  \textrm{ s}
\end{equation}
However, Chen et al. obtained a linear scaling for DNA 1.0-16.5 $\mu$m in length \cite{chen2004probing}:
\begin{equation}
\tau_\mathrm{trans} \approx 1.0 \times 10^{-4} L^1 \textrm{ s}
\end{equation}
Both of these results are for an applied voltage of 120 mV, which is a typical value \cite{smeets2006salt,tabard2007noise,Mihovilovic2013statistics,meller2001voltage}. 
Voltages as low as 20 mV \cite{fologea2005slowing} and as high as $> 800$ mV \cite{chen2004probing} have also been used. Using the $\tau_\mathrm{trans}$ forms given above, we obtain
\begin{eqnarray}
P_\mathrm{Storm} & \approx & 160 ~ L^{0.54}\\
P_\mathrm{Chen} & \approx & 90 ~ L^{0.81}
\label{pec_ex}
\end{eqnarray}
The scaling exponents (for $N$ or $L$) found in the expressions for $P_\mathrm{sim}$, $P_\mathrm{Chen}$ and $P_\mathrm{Storm}$ are slightly different. 
One significant deviation between the simulations and experiments is that the simulations do not include hydrodynamic effects.
In spite of these differences, the scaling is rather similar and thus achieving the correct $P_t$ at one polymer size will result in essentially the correct $P_t$ for a fairly range of sizes, as we shall see.

The last step before we can compare simulation and experimental P\'{e}clet numbers is the selection of a length scale, i.e., we need to choose the monomer bead size ($\sigma$). 
For dsDNA we can set the bead diameter to be 10~nm, a convenient value that is only slightly larger than estimates for the effective width of DNA (5-10~nm) \cite{rybenkov1997effect, wang2011simulation, dai2013revisiting}. 
Since the simulation pore radius is set to be $\sigma$, $\sigma=10$~nm implies a pore with a diameter of 20~nm. 
This compares well to the values of 10~nm in Storm et al. \cite{storm2005fast} and 15~nm in Chen et al. \cite{chen2004probing}. 
With $\sigma=10$~nm, a polymer of 100 beads corresponds to a DNA fragment with a contour length of 1 $\mu $m.
This corresponds to the lower bound of the Chen data cited above, but if we consider doing a scaling experiment and extending $N$ up to 400,
then the corresponding lengths would encompass the lowest two or three data points of both the Chen and Storm data.

\begin{figure}[h]
 	\centering
	\includegraphics[width=0.50\textwidth]{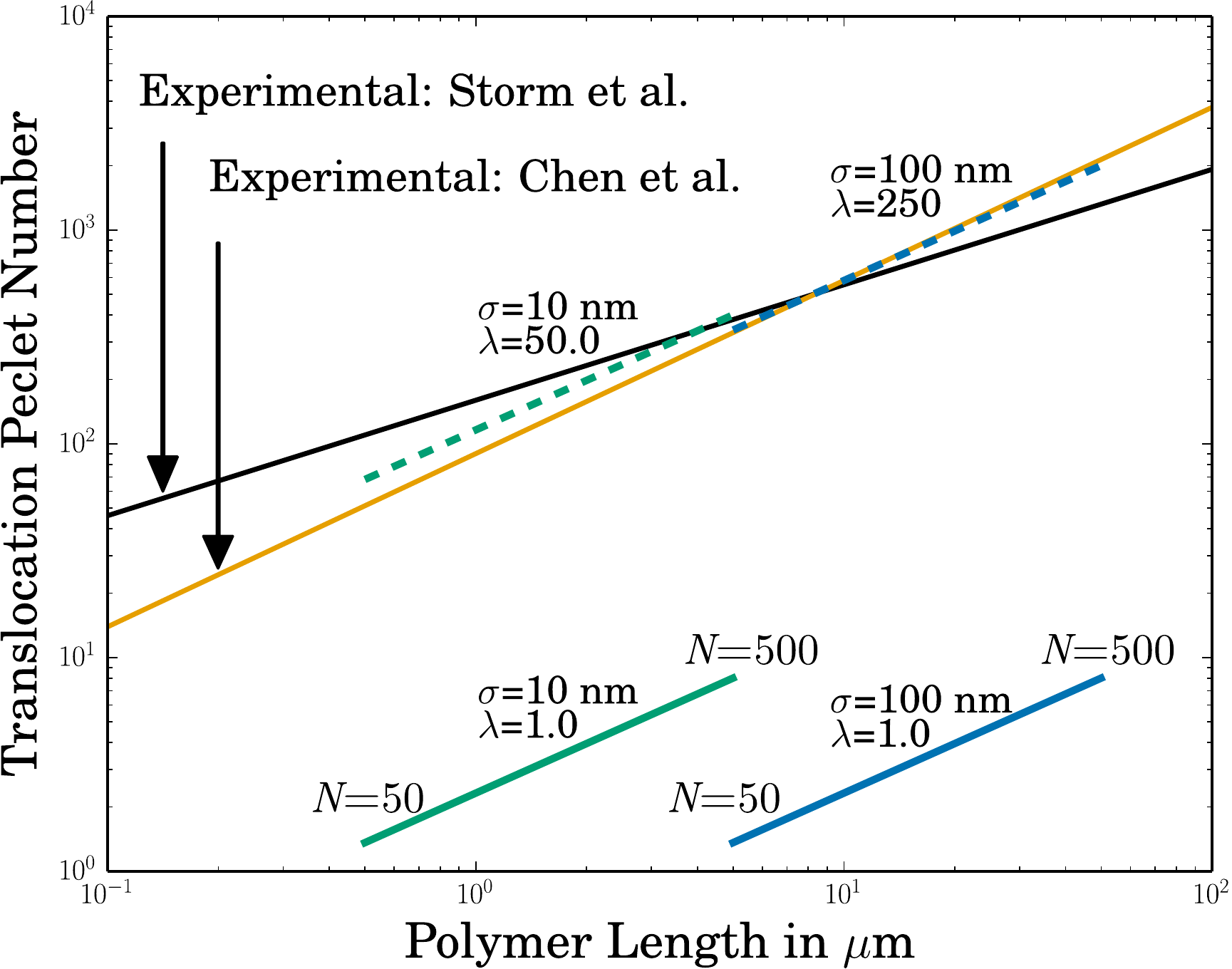} 
	\caption{Experimental and simulation P\'{e}clet numbers as a function of polymer length.
	The experimental $P_t$ based on the results of Storm et al. \cite{storm2005fast} and Chen et al. \cite{chen2004probing} have slightly different scalings with respect to polymer length.
	The simulation $P_t$ are shown for two different length scales:
	$\sigma$=10~nm is shown in green and $\sigma$=100~nm is shown in blue.
	For the former, $\lambda$=50 brings the simulation and experimental $P_t$ into good agreement;
	for the latter, $\lambda$=100 is required.}
	\label{fig:peclet}
\end{figure}

The P\'{e}clet numbers from Eq. \ref{pec_sim} and  \ref{pec_ex} are shown in Fig. \ref{fig:peclet}.
For $\sigma$=10~nm and $\lambda=1.0$, $P_\mathrm{sim}$ is significantly lower than the estimate from experiments for a given contour length. 
Hence, this CG model yields an artificially low P\'{e}clet number: diffusion is much greater than it should be.
To correct this, we can choose $\lambda=50$: this brings the data for $P_t$ from simulation and experiments into much closer agreement. 
This indicates that to approximate the experimental balance between diffusion and drift, the diffusivity of the polymer should be lowered to 1/50 of its ``default" simulation value.

To compare these values to typical CG setups, refer again to the list of parameters used in various simulation studies as shown in Table I. 
Most studies take $k_\mathrm{B}T \gtrsim \epsilon$, a force value of 1--10 $\epsilon/\sigma$, and polymers on the order of $N=100$ beads.
For these typical values we obtain $P_t \approx 1$--$10$; this Peclet range is indicated by a shaded area in the plots of the results section (Figs. \ref{fig:tau}, \ref{fig:prob} and  \ref{fig:std}). In contrast to this, we find that $P_t \approx 50$ would be required to model experimental dynamics.
As will be shown, the discrepancy grows for greater amounts of coarse-graining (i.e., $\sigma$ corresponding to larger length scales).
From this, we suggest that diffusion effects are too prominent in the majority of CG simulation setups for studies of the translocation of dsDNA.

Examining the figures in the results section, $\lambda=50$ lies within the driven translocation regime.
Hence, translocation is primarily driven and, contrary to most coarse-grained setups, diffusion is a smaller effect.
Further, most simulations are being performed in the transition regime where the behaviour of the probability, $\tau$, and $\sigma_{st}$ are all changing.
This could greatly complicate the interpretation of the simulation results and hinder the experimental relevance.

Further, as most studies employ a freely jointed chain (FJC) model, it is unclear that they are modelling at the precision of $\sigma$=10~nm since a FJC model of dsDNA implies that $\sigma$ is at least a Kuhn length (i.e., 100~nm).
Taking $\sigma$=100~nm would have the added advantage of modelling longer DNA strands for a fixed value of $N$.
For $\sigma$=10~nm, polymer lengths of $N=100$ or 200 beads (as typically studied) correspond to DNA strands 1 or 2 $\mu$m in length. 
On the other hand, $\lambda$-DNA, 
which is often used in translocation experiments \cite{Mihovilovic2013statistics, smeets2006salt, Storm2005translocation}, 
has a length around 16~$\mu$m and is thus about an order of magnitude longer and thus one might choose $\sigma$=100~nm, or simulate longer chains with higher $N$. 
However, this choice also implies a nanopore with a radius of 100~nm. 
While experiments are performed with pores as large as this, most research focuses on tighter pores, particularly so for applications such as sizing and sequencing DNA.

For this reason, $\sigma$=100~nm introduces complications to the modelling that are avoided by choosing $\sigma$= 10~nm.
However, since it is instructive to examine the $\lambda$ required to match results for the coarser modelling, the $\sigma$=100~nm results are also shown in Fig. \ref{fig:peclet}. 
For this case, 
the $\lambda$=1.0 data underestimates the translocation P\'{e}clet number to an even greater degree 
and to match the numbers, a coarse-graining factor of $\lambda=250$ is needed.
\subsection{Single Stranded DNA}
Translocation of ssDNA presents a slightly more complicated scenario than dsDNA.
To ensure single-file translocation of ssDNA the nanopores need to be smaller than the ones used for dsDNA.
Although most people turn to biological protein nanopores for this, we will focus on work using solid-state nanopores \cite{briggs2014, heng2004sizing, venta2013differentiation, fologea2005detecting,kowalczyk2010,fyta2011translocation,wanunu2008dna} as these should have lower interactions with the ssDNA molecule and are closer to the `hole-in-wall' geometries studied in typical simulations.
To complicate matters, ssDNA will tend to self-hybridize and form hairpins which fundamentally changes the nature of the translocation process \cite{kowalczyk2010}.
This could make ssDNA a bad candidate for the generic model polymer studied in the CG simulations.

Many translocation studies of ssDNA have been restricted to very short molecules (i.e., below than 50  bases) \cite{heng2004sizing, venta2013differentiation}.
However, in 2005 Fologea et al. achieved the translocation of a 3~kb strand through a nanopore \cite{fologea2005detecting}.
Using a pH of 13 to prevent hairpin formation from self-hybridization, they measured a translocation time of $\tau$=120~$\mu$s using a voltage of 120 mV.

Although factors such as pH and finite size effects complicate matters, we now develop a rough estimate for the P\'{e}clet number for this experiment.
Modelling this as a freely jointed chain, we can set $\sigma$ to be the Kuhn length, 7~nm (note this is much more flexible than dsDNA) \cite{tinland1997persistence},
and thus the 3~kb strand corresponds to $\approx$ 190 beads -- a value that is well within the range of CG simulations  \cite{linna2012event}.
Using the Kratky-Porod (worm-like chain) model, we estimate a radius of gyration of 38~nm for this strand.
From Tinland et al. \cite{tinland1997persistence}, we estimate the diffusion coefficient for a 3~kb strand to be $D \approx 4$ $\mu$m$^2$/s.
Using all of this in Eq. \ref{good_pec}, we find
\begin{equation}
P_\mathrm{exp} \approx 2.
\end{equation}
Hence, while the $P_t$ in typical CG setups is too low for dsDNA, it is about right for ssDNA with a length on the order of 3~kb (in agreement with the simulations of Linna and Kaski \cite{linna2012event}).
Recall that this is a relatively long strand of ssDNA for translocation studies.
In the studies performed with shorter lengths \cite{venta2013differentiation,heng2004sizing},
$P_t$ will be even lower -- indicating that diffusion is even more prominent.
This result indicates that while translocation of dsDNA is certainly a non-equilibrium process,
it may indeed be a quasi-static process for short ssDNA strands
(here quasi-static indicates that the relaxation time of the polymer is much shorter than the translocation time such that the polymer is essentially relaxed at each stage of translocation) \cite{dehaanquasi}.
This approximation was assumed in the early theoretical work predicting scaling laws \cite{muthu1999,sung1996}
and has been used in many theoretical and simulation studies since.
While generally regarded as an unphysical limit, the above results indicate that a quasi-static process may be achievable for short ssDNA strands.

\subsection{Rod-like Viruses}

One of the authors has also used the P\'{e}clet number approach to match simulation studies to experimental results for the translocation of rigid fd viruses through nanopores \cite{mcmullen2014}.
With a persistence length $l_\mathrm{p}$ that is greater than three times the contour length $L$, the viruses are rod-like and thus $L$ is a more convenient length-scale in Eq.~\ref{good_pec} 
than the radius of gyration.
In this study, $\lambda$ was varied between 1 and 12 for comparison to experimentally relevant voltages.
As the results were found to both qualitatively and quantitatively depend on the strength of the applied field,
matching the simulations to the experimental conditions was crucial for having the simulations shed light on the experimental results.
The P\'{e}clet number approach allowed for a very straightforward matching of diffusion-drift effects using measurements that were easily obtainable in both experiments and simulations.

\section{Conclusions}
Many simulation studies of  polymer translocation have employed a coarse-grained methodology.
Although DNA translocation is frequently cited as a motivating application,
the CG models are not often matched to experimental conditions.
In this work, we propose a method for tuning one crucial aspect, the balance between drift and diffusion, to experimental conditions via a P\'eclet number for translocation.

Using this definition, we have demonstrated that the drift-diffusion balance in coarse-grained simulations can dramatically alter the physics of translocation.
Mapping out three regimes (low, transition, and high $\lambda$ values), different -- and sometimes opposing results -- are obtained for fundamental translocation measurements.
For both in-pore and pulling forces,
the probability of translocation is observed to go from very low values at low $\lambda$ to 100\% translocation rates at high $\lambda$.
More surprisingly, the translocation time, $\tau$, is found to vary non-monotonically across the three regimes.
Finally, the variance in the translocation time is also found to be sensitive to the drift-diffusion balance.

Simulating in a regime inappropriate for comparison to experiments will thus have several consequences.
First, the translocation probability will be incorrect.
As most coarse-grained setups tend to over emphasize diffusion, the probability of retraction to $cis$ (i.e., a failed event) has likely been over-stated.
Second, if we consider simulations with a wide range of polymer lengths, it is possible that short polymers will fall in one regime while long polymers fall in another.
Not only does this complicate the calculation of scaling exponents,
but agreement between simulation and experiment could be compromised if the span of differing regimes is not the same.
The non-monotonic behaviour of $\tau$ with $\lambda$ will thus add a complicating factor to the determination of scaling laws;
the details of regime dependence invalidate the generality implied by scaling laws.
This may help to explain the persistent disagreement between simulation and experimental scaling exponents. 

Third, the disparate results between the two force models for $\sigma_{st}/\tau$ at high $\lambda$ values demonstrates the impact that the non-equilibrium, crowding effects can have on translocation.
For the in-pore force, $\sigma_{st}/\tau$ is non-monotonic and begins to increase with $\lambda$ while for the pulling force, $\sigma_{st}/\tau$ continues to decay.
Again, coarse-grained simulations typically over-emphasize diffusion and thus these crowding effects are not significant.
Hence, the results may be significantly different at proper $\lambda$ values where crowding plays a large role in both the $\sigma_{st}/\tau$ and $\tau$ values.

We have examined several different experimental translocation scenarios and estimated the P\'eclet $P_t$ and corresponding coarse-graining parameter $\lambda$ for each case.
Most significantly, we find that for translocation of dsDNA, $P_t$ for typical simulations is anywhere from 5--50 times lower than for experimental studies.
While recent progress using a tension-propagation model has united many simulation results \cite{ikonen2012b}, 
the discrepancy with experiments remains \cite{storm2005fast, chen2004probing, luo2009driven,  bhattacharya2009scaling}.
This may be at least partially explained by the fact that the crowding effects on the exit ($trans$ side) have not been implemented in detail in this approach.
Likewise, performing CG simulations at the proper $P_t$ such that crowding effects have the correct impact may bring simulations and experiments in closer agreement.

Hence, obtaining the correct balance of drift to diffusion is crucial for modelling translocation and can be achieved this by using a P\'{e}clet number for translocation that is easily calculated for both experimental and simulation conditions.
Achieving the same P\'eclet number in the simulations is accomplished by matching length-scales via the coarse-graining $\lambda$ parameter.
While the examples herein outline the effect that $P_t$ has on the results, there are many more translocation scenarios where the impact could be felt. 

\section{Acknowledgements}

Simulations were performed using the ESPResSo package \cite{Espresso} on the SHARCNET computer system (www.sharcnet.ca) using VMD \cite{vmd} for visualization.
Graphics produced with Matplotlib \cite{matplotlib}.
This work was funded by NSERC and the University of Ottawa.

\bibliography{Peclet.bib}


\end{document}